\documentclass[sigconf]{acmart}

\usepackage[english]{babel}
\usepackage{blindtext}
\usepackage{makecell}
\usepackage{booktabs}
\usepackage{graphicx}
\usepackage{hyperref}
\usepackage{indentfirst}
\usepackage{float}

\usepackage{listings}
\usepackage{xcolor}

\lstdefinelanguage{Rust}{
  keywords={
    abstract, as, async, await, become, box, break, const, continue, crate,
    do, dyn, else, enum, extern, false, final, fn, for, if, impl, in, let, loop,
    macro, match, mod, move, mut, override, priv, pub, ref, return, self, Self,
    static, struct, super, trait, true, try, type, typeof, unsafe, unsized, use,
    virtual, where, while, yield, func, mapping, var, program, Ok
  },
  sensitive=true,
  comment=[l]{//},
  morecomment=[s]{/*}{*/},
  morestring=[b]",
  morestring=[b]',
}

\usepackage{titlesec}
\titleformat{\subsubsection}[block]{\bfseries}{\thesubsubsection}{1em}{} 
\titlespacing{\subsubsection}{0pt}{\baselineskip}{1em}

\usepackage{enumitem}
\setlist{itemsep=0.5em}

\renewcommand\footnotetextcopyrightpermission[1]{} 
\setcopyright{none}

\acmDOI{}

\acmISBN{}

\acmPrice{}

\begin{document}

\title{zkSDK: Streamlining zero-knowledge proof development through automated trace-driven ZK-backend selection}

\author{
{\rm William Law}\\
University of Waterloo\\
william.law@uwaterloo.ca
}

\renewcommand{\shortauthors}{William Law}

\begin{abstract}
The rapid advancement of creating Zero-Knowledge (ZK) programs has led to the development of numerous tools designed to support developers. Popular options include being able to write in general-purpose programming languages like Rust from Risc Zero. Other languages exist like Circom, Libsnark, and Cairo. However, developers entering the ZK space are faced with many different ZK backends to choose from, leading to a steep learning curve and a fragmented developer experience across different platforms. As a result, many developers tend to select a single ZK backend and remain tied to it. This thesis introduces zkSDK, a modular framework that streamlines ZK application development by abstracting the backend complexities. At the core of zkSDK is Presto, a custom Python-like programming language that enables the profiling and analysis of a program to assess its computational workload intensity. Combined with user-defined criteria, zkSDK employs a dynamic selection algorithm to automatically choose the optimal ZK-proving backend. Through an in-depth analysis and evaluation of real-world workloads, we demonstrate that zkSDK effectively selects the best-suited backend from a set of supported ZK backends, delivering a seamless and user-friendly development experience.
\end{abstract}

\maketitle
\title{zkSDK}

\section{Introduction}

\subsection{Problem}

Zero-knowledge programs are a way to prove that you know something without revealing what it is. It solves the problem of verifying the validity of information or computations without revealing the underlying data. This ensures privacy, which has many use cases, from personal to business and legal applications.

This thesis investigates the development of a Dynamic Backend Selection Mechanism (DBSM) designed to optimize the performance of the ZK program by selecting the most suitable proving backend based on user-defined requirements and computational traces. Users can specify performance objectives such as proof generation time, gas-efficient proof verification, or the need for hardware acceleration, allowing the DBSM to make informed backend selections.

In order to supply the DBSM with the computational traces of the program workload, we needed to develop Presto, a custom programming language that has a minimal feature set. These traces are important for understanding where the program spends most of its time. This approach allows our compiler to have precise control over the structure of the Abstract Syntax Tree (AST), enabling it to profile and analyze the computational workload of a program before passing it to the DBSM. Additionally, the compiler retains the full context of the AST during code generation, ensuring that it can accurately translate Presto code into the code corresponding to the selected ZK backend. It should be noted that it is still possible to do this without a custom language. Together, this aims to address the gaps in traditional ZK development where users are required to choose and manually configure a proving backend from many different options.

\subsection{Motivation}

The advancements of generating ZK proofs (ZKPs) made it possible to write ZK applications, opening the opportunity for different ZK backends to help developers write these programs. However, each backend comes with its trade-offs in performance and usability. For example, this could be developers needing to learn a new programming language, ranging from Rust to Circom, or learning a new programming paradigm like writing ZK circuits. With the promising technology of ZKPs, the fragmented ecosystem makes onboarding difficult for new developers, limiting accessibility.

This thesis aims to streamline and standardize the different backends available to write ZKPs. Instead of having developers learn multiple languages and frameworks, developers using the zkSDK will only need to learn and interface with Presto. Given that many developers are familiar with Python, Presto offers a similar experience, reducing the onboarding time while preserving the expressiveness needed for writing ZKPs. Although developers can still explore these backends independently, Presto reduces the barrier to entry, making it easier for those who want to start writing ZK code quickly without the steep learning curve.

\subsection{Contributions}

In this thesis, we introduce the zkSDK that streamlines ZK development through automated trace-driven ZK backend selection. The main contributions include:

\begin{itemize}
  \item \textbf{Assessing ZK backend performance:} To understand the performance and trade-offs of different ZK backends, we benchmarked the Risc Zero zero-knowledge virtual machine (zkVM) and Consensys Gnark as two initial ZK backends to support. The benchmarks consisted of tasks across integer operations, ECDSA signature verification, computing ZK-friendly hashes (i.e., MiMC), and general-purpose cryptographic hashes (i.e., SHA-2 and SHA-3). These tasks were measured based on proof generation time and gas usage to verify proofs on Ethereum. This data serves as input to the DBSM.
  \item \textbf{Presto language:} A Python-like syntax that provides similar expressiveness as other languages used to write ZK programs. The zkSDK utilizes \href{https://pest.rs/}{Pest}, a Rust-based general-purpose parser, along with custom AST parsing logic to translate Presto code into an AST data structure in Rust.
  \item \textbf{Profiling and analysis:} The zkSDK introduces the Interpreter package to perform the computational analysis of the Presto AST. It is responsible for traversing the AST and identifying certain operations that contribute to the computational intensity of proving ZK applications. These statistics are collected and used by the DBSM to determine the most suitable backend for translating the Presto code. 
  \item \textbf{Code Generation:} To enable multi-backend support, the zkSDK Codegen package is responsible for handling the translation of the Presto code to the appropriate ZK backend code that the DBSM optimally selects.
  \item \textbf{Selection algorithm:} The zkSDK DynamicSelect package handles the input of user preferences and Interpreter results by converting this data into Rust \href{https://docs.rs/nalgebra/latest/nalgebra/}{nalgebra} matrices. The DBSM normalizes these inputs and performs matrix multiplication to select the minimum value corresponding to one of the ZK backends.
\end{itemize}
\section{Background}

\subsection{Zero-Knowledge Proofs}

The key components involved in a ZKP are the problem, the prover, the verifier, and a degree of randomness. These are some useful information to keep in mind about ZK \cite{chainlinkzk}:

\begin{itemize}
  \item \textbf{Completeness:} If a statement is true, then an honest verifier can be convinced by an honest prover that they possess knowledge about the correct input.
  \item \textbf{Soundness:} If a statement is false, then no dishonest prover can unilaterally convince an honest verifier that they possess knowledge about the correct input.
  \item \textbf{Zero-knowledge:} If the state is true, then the verifier learns nothing more from the prover other than the statement is true.
  \item \textbf{ZK Circuit:} An arithmetic circuit over a finite field that encodes a computation as a set of constraints (R1CS or PLONK-style gates) to enable zero-knowledge proof generation and verification
\end{itemize}

To generate these ZKPs, there are different approaches that make trade-offs between the proof size, proving time, and verification time. Some of the common ones include:

\begin{itemize}
  \item \textbf{SNARKs:} A Succinct Non-interactive Argument of Knowledge provides the ability for one party to prove to another that they know a secret without revealing the secret itself. \cite{chainlinksnark}
  \item \textbf{STARKs:} A Scalable Transparent Argument of Knowledge is an alternative to SNARKs. In the application of blockchains, STARKs increases scalability by allowing the proofs to be generated off-chain, which can be posted later onchain and verified by a STARK verifier. \cite{chainlinksnark}
  \item \textbf{PLONK:} A proving system that offers a universal and efficient zero-knowledge proof system that allows proving statements succinctly and verifiably with a single setup. \cite{plonk}
  \item \textbf{Groth16:} A proving system that produces constant-size proofs and fast verification but requires a trusted setup for each circuit.
\end{itemize}

\subsection{Zero-Knowledge Backends}

\subsubsection{Risc Zero zkVM}

The Risc Zero (risc0) zkVM is a RISC-V virtual machine that enables developers to write ZK applications in mature languages like Rust and C++. The risc0 zkVM can prove the correct execution of arbitrary code without having to build a ZK Circuit \cite{risczero}. The zkVM contains three main components:

\begin{itemize}
  \item \textbf{Guest:} This is where most of the core logic for a ZK application goes and is the code that will be executed and proven by the zkVM. The guest program can write public outputs to a journal, which will be included in the receipt.
  \item \textbf{Host:} The machine running the zkVM that constructs the ZKP. The host manages the inputs and outputs of the guest program. The zkVM runs the Executor that executes the Guest program followed by the Prover, which generates a receipt.
  \item \textbf{Receipts:} Attests to the valid execution of a guest program. The receipt contains fields like the journal, imageID, and seal. The imageID is a cryptographic identifier that indicates the entry point in the guest program. The seal attests to the correct execution of the guest program \cite{risczeroreceipt}.
\end{itemize}

\begin{figure}[h]
\centering
\includegraphics[width=0.9\columnwidth]{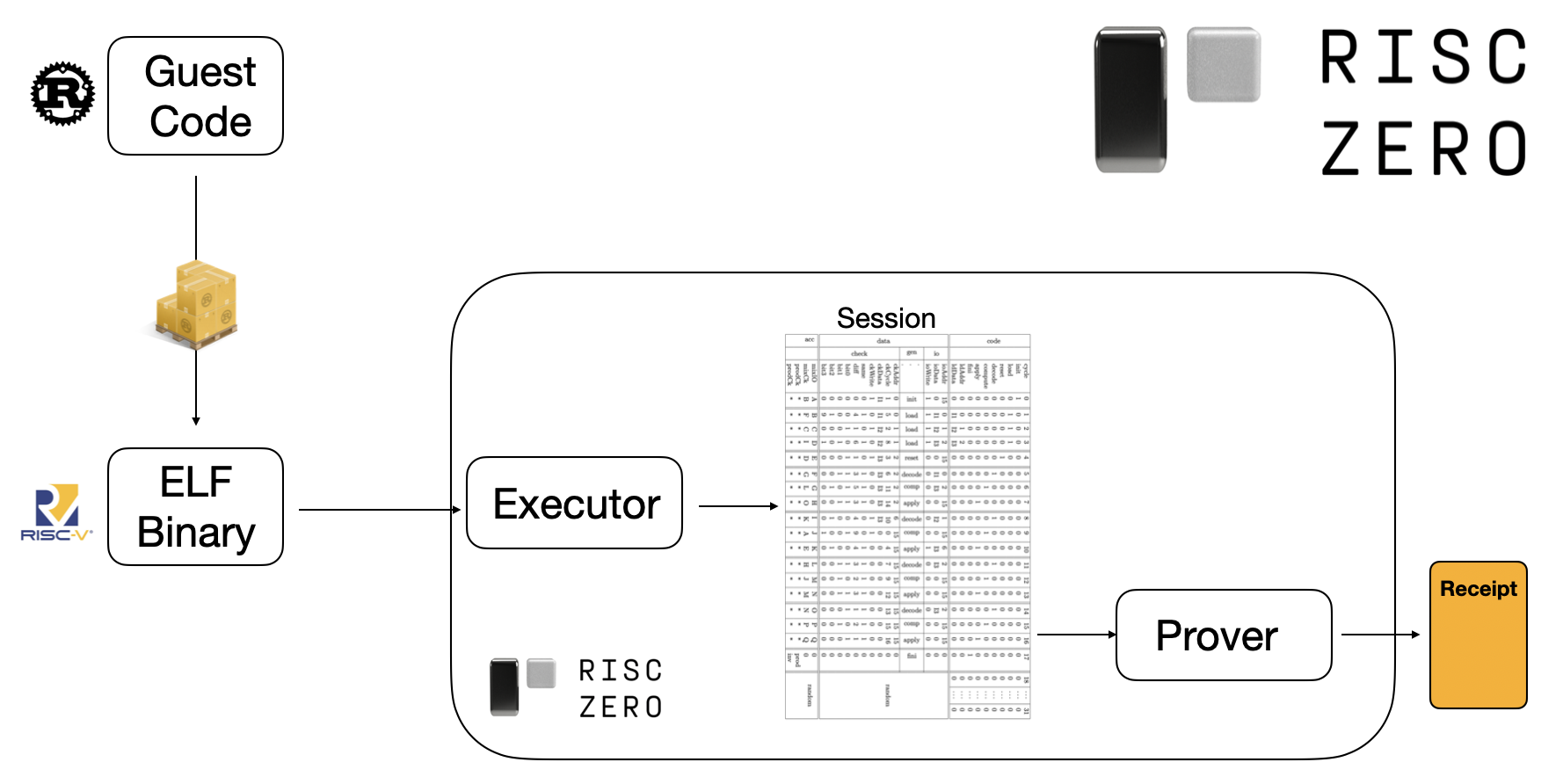}
\caption{Diagram of a Risc Zero codebase \cite{risczerozvm}}
\label{fig:risc-zero-codebase}
\end{figure}

The zkVM provides tools for generating proofs in development mode, which will show the total cycles of the program. Cycles are the smallest unit of compute in the zkVM, similar to a clock cycle on a physical CPU.

When locally proving ZK code, it is important to understand that the generated receipt is a Succinct Receipt. This receipt is produced by zkVM, which splits the program into Segments and proves each Segment individually before combining them into a Succinct Receipt. To generate Groth16 Receipts, there is a specific requirement to use a Linux x86 machine or a remote proving service like Bonsai \cite{risczeroproving}. This is because the circuits in the risc0 zkVM are compiled to Circom, which produces x86 assembly code.

\subsubsection{Consensys Gnark}

Consensys created a SNARK library called Gnark that enables developers to easily write their own circuits through its high-level API. These circuits are written in Golang, where all interaction is done via Gnark's custom constraint variables, called \texttt{frontend.\allowbreak Variable}. The developers utilize the Frontend package to perform operations, comparisons, and compile their circuit. 

Once the circuit is compiled, the Backend package is used to create and verify ZKPs. To do this, Gnark offers a \texttt{Setup}, \texttt{Prove}, and \texttt{Verify} interface for each SNARK schema it supports, which currently includes Groth16 and Plonk. The \texttt{Prove} and \texttt{Verify} take a witness as input. Witnesses contains inputs that are needed by the circuit. For verification, the witness only includes public inputs, referred as a Public Witness. For proving, a Full Witness includes both public and secret inputs \cite{gnarkwit}.

\section{System Architecture}

\subsection{Overview}

At a high-level, the end-to-end workflow is as follows. A developer writes a ZK program in Presto. When the developer is ready to compile their code, they pass in a list of weights representing the predefined supported user preferences. The Presto file will be passed to the compiler and transformed into an AST in Rust. The zkSDK compiler consists of three components: AST Parser, Interpreter, and Codegen. The Interpreter will traverse the AST to determine the computational intensity of the program. Along with the user preferences, the DBSM will select an appropriate ZK backend (risc0 or Gnark) and the Codegen will produce the corresponding backend code. 

\begin{figure*}[h]
\centering
\includegraphics[width=0.9\textwidth]{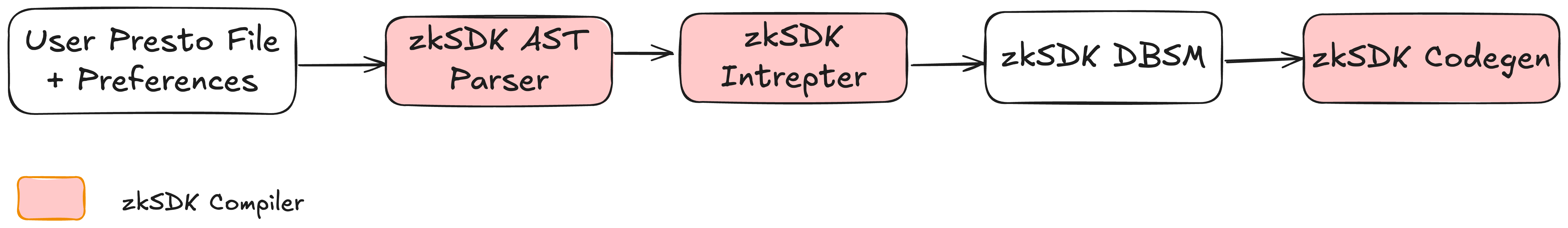}
\caption{zkSDK end-to-end workflow}
\label{fig:zksdk-workflow}
\end{figure*}

\subsection{Components}

\subsubsection{Benchmarks}

We first performed benchmarks on both risc0 and Gnark on the time to generate proofs and the gas cost to verify them on the Ethereum blockchain. This gives us a sense of the tradeoffs and strengths of the different backends. It is important to highlight that these numbers will be represented as a matrix to be used by the DBSM to calculate the optimal backend.

\begin{table*}
\centering
\caption{Integer Operations (Add y += x**3 + y + 5 100,000 times, where initially y=x, x=1024)}
\begin{tabular}{lllrrrr}
\toprule
Backend & Prover & Curve & Time (s) & Total Cycles & No. Constraints & Gas Usage \\
\midrule
Risc Zero & - & - & 26.576 & 1,310,720 & - & - \\
GNARK & Groth16 & BN254 & 8.688 & - & 200,335 & 194,955 \\
GNARK & Groth16 & BLS12-381 & 13.606 & - & 200,335 & - \\
GNARK & Plonk & BN254 & 4.091 & - & 100,002 & 262,869 \\
GNARK & Plonk & BLS12-381 & 6.366 & - & 100,002 & - \\
Risc Zero & Groth16 & BN254 & - & - & - & 229,894 \\
\bottomrule
\end{tabular}
\label{tab:int-ops}
\end{table*}

\begin{table*}
\centering
\caption{ECDSA (Proving the message "Hello World")}
\begin{tabular}{lllrrrr}
\toprule
Backend & Prover & Curve & Time (s) & No. Constraints & Total Cycles & Gas \\
\midrule
Risc Zero & - & - & 267.000 & - & 13,631,488 & - \\
GNARK & Groth16 & BN254 & 42.679 & 293,814 & - & 317,105 \\
GNARK & Groth16 & BLS12-381 & 66.661 & 293,814 & - & - \\
GNARK & Plonk & BN254 & 63.697 & 1,135,876 & - & 281,458 \\
GNARK & Plonk & BLS12-381 & 92.562 & 1,135,876 & - & - \\
Risc Zero & Groth16 & BN254 & - & - & - & 230,618 \\
\bottomrule
\end{tabular}
\label{tab:ecdsa}
\end{table*}

\begin{table*}
\centering
\caption{MiMC (Proving 100 MiMC Hashes)}
\begin{tabular}{lllrrrr}
\toprule
Backend & Prover & Curve & Time (ms) & No. Constraints & Total Cycles & Gas \\
\midrule
Risc Zero & - & - & 1,576.162 & - & 65,536 & - \\
GNARK & Groth16 & BN254 & 2,446.640 & 33,001 & - & 194,976 \\
GNARK & Groth16 & BLS12-381 & 3,737.577 & 33,001 & - & - \\
GNARK & Plonk & BN254 & 2,144.795 & 44,299 & - & 262,847 \\
GNARK & Plonk & BLS12-381 & 3,232.549 & 44,699 & - & - \\
Risc Zero & Groth16 & BN254 & - & - & - & 229,872 \\
\bottomrule
\end{tabular}
\label{tab:mimc}
\end{table*}

\begin{table*}
\centering
\caption{SHA2 Benchmarks}
\begin{tabular}{llllrrrr}
\toprule
Backend & Prover & Curve & Job Size & Mean (s) & Std Dev (s) & No. Constraints & Gas \\
\midrule
Risc Zero & - & - & 1024 & 2.194 & 0.395 & - & - \\
Risc Zero & - & - & 2048 & 5.000 & 0.000 & - & - \\
Risc Zero & - & - & 4096 & 10.429 & 0.728 & - & - \\
Risc Zero & - & - & 8192 & 23.000 & 1.633 & - & - \\
GNARK & Groth16 & BN254 & 1024 & 5.980 & 0.450 & 599,888 & 317,083 \\
GNARK & Groth16 & BN254 & 2048 & 12.310 & 0.340 & 1,041,927 & - \\
GNARK & Groth16 & BN254 & 4096 & 33.460 & 0.470 & 1,926,006 & - \\
GNARK & Groth16 & BN254 & 8192 & 104.430 & 0.800 & 3,694,163 & - \\
GNARK & Groth16 & BLS12-381 & 1024 & 11.830 & 0.000 & 599,888 & - \\
GNARK & Groth16 & BLS12-381 & 2048 & 25.450 & 0.000 & 1,041,927 & - \\
GNARK & Groth16 & BLS12-381 & 4096 & 71.960 & 0.000 & 1,926,006 & - \\
GNARK & Groth16 & BLS12-381 & 8192 & 198.660 & 0.710 & 3,694,163 & - \\
GNARK & Plonk & BN254 & 1024 & 76.800 & 0.820 & 2,182,838 & 281,502 \\
GNARK & Plonk & BN254 & 2048 & 55.270 & 0.940 & 3,747,574 & - \\
GNARK & Plonk & BN254 & 4096 & 198.090 & 0.000 & 6,877,046 & - \\
GNARK & Plonk & BN254 & 8192 & 451.220 & 8.500 & 13,135,990 & - \\
GNARK & Plonk & BLS12-381 & 1024 & 128.270 & 2.000 & 2,182,838 & - \\
GNARK & Plonk & BLS12-381 & 2048 & 72.360 & 2.000 & 3,747,574 & - \\
GNARK & Plonk & BLS12-381 & 4096 & 266.010 & 6.500 & 6,877,046 & - \\
GNARK & Plonk & BLS12-381 & 8192 & 593.440 & 22.500 & 13,135,990 & - \\
Risc Zero & Groth16 & BN254 & - & - & - & - & 230,391 \\
\bottomrule
\end{tabular}
\label{tab:sha2}
\end{table*}

\begin{table*}
\centering
\caption{SHA3 (Keccak 256) Benchmarks}
\begin{tabular}{llllrrrr}
\toprule
Backend & Prover & Curve & Job Size & Mean (s) & Std Dev (s) & No. Constraints & Gas \\
\midrule
Risc Zero & - & - & 1024 & 9.714 & 0.700 & - & - \\
Risc Zero & - & - & 2048 & 21.667 & 0.471 & - & - \\
Risc Zero & - & - & 4096 & 26.000 & 1.414 & - & - \\
Risc Zero & - & - & 8192 & 49.000 & 0.000 & - & - \\
GNARK & Groth16 & BN254 & 1024 & 7.190 & 0.000 & 613,183 & 317,105 \\
GNARK & Groth16 & BN254 & 2048 & 18.920 & 0.390 & 1,094,682 & - \\
GNARK & Groth16 & BN254 & 4096 & 41.060 & 0.500 & 1,997,495 & - \\
GNARK & Groth16 & BN254 & 8192 & 117.860 & 0.630 & 3,803,120 & - \\
GNARK & Groth16 & BLS12-381 & 1024 & 15.610 & 0.470 & 613,183 & - \\
GNARK & Groth16 & BLS12-381 & 2048 & 39.490 & 0.480 & 1,094,682 & - \\
GNARK & Groth16 & BLS12-381 & 4096 & 84.180 & 0.000 & 1,997,495 & - \\
GNARK & Groth16 & BLS12-381 & 8192 & 231.950 & 0.830 & 3,803,120 & - \\
GNARK & Plonk & BN254 & 1024 & 80.060 & 0.940 & 2,537,563 & 281,458 \\
GNARK & Plonk & BN254 & 2048 & 194.280 & 2.500 & 4,550,531 & - \\
GNARK & Plonk & BN254 & 4096 & 125.890 & 4.500 & 8,324,974 & - \\
GNARK & Plonk & BN254 & 8192 & 474.970 & 16.500 & 15,873,860 & - \\
GNARK & Plonk & BLS12-381 & 1024 & 129.280 & 1.000 & 2,537,563 & - \\
GNARK & Plonk & BLS12-381 & 2048 & 253.930 & 1.000 & 4,550,531 & - \\
Risc Zero & Groth16 & BN254 & - & - & - & - & 230,458 \\
\bottomrule
\end{tabular}
\label{tab:sha3}
\end{table*}

These benchmarks were done on a Macbook M1 Pro \allowbreak(arm64) with 32GB of RAM on MacOS 14.6.1. The target bench time was 60 seconds. For the full setup, see Section 5.1.

It is important to highlight that the Risc Zero Proof Verification tests were failing due to a reason that we could not figure out. However, since Groth16 proof sizes are constant, and knowing that the Solidity revert happened at the end of the test, the reported gas usage gives us a close enough approximation. It is also important to note that these Risc Zero Proof Verification tests were generated using Bonsai, Risc Zero's remote proving service, as STARK-to-SNARK was not supported on arm64. 

\subsubsection{Grammar}

To build Presto, the zkSDK uses Pest, a general-purpose parser, to seamlessly create a simple grammar. This grammar is written into a \texttt{.pest} file. Presto is a statically typed language that supports custom function definitions, variable declarations, mapping structures, binary and integer operations, byte manipulation, conditionals, loops, and structs. It also includes a set of precompiles, such as ECDSA, SHA256, Keccak256, and MiMC. 

A Presto file starts with the \texttt{program} keyword and the name of the file. The body of the program is represented as \texttt{statements}, which is a list containing either a \texttt{statement} or optional semicolon statement (\texttt{oscs}). A \texttt{statement} is a variable declaration, mapping, or custom function call and requires a semicolon at the end of the line. Whereas a \texttt{oscs}, would be an if-statement, struct declaration, function declaration, and a for-loop that does not require a semicolon at the end of the line. See Section 5.2 for Presto examples.

\subsubsection{AST Parser}

Although Pest outputs an AST object, we built additional parsing logic around this AST object to convert it into a Rust object type that is easily reusable in the later stages of the compiler. We created a trait called \texttt{try\_parse} \cite{riftv2} that will recursively traverse the elements of \texttt{statement} and \texttt{oscs} and translate the Pest-generated AST into reusable Rust objects. 

\begin{lstlisting}[breaklines]{rust}
trait PrestoNode<T> {
    fn parse(pair: Pair<Rule>, i: usize) -> Result<T>;
    fn ast_type() -> Vec<Rule>;
}

fn try_parse<F>(pair: Pair<Rule>) -> Result<F>
where
    F: PrestoNode<F>,
{
    let r = pair.as_rule();
    for (i, rtype) in F::ast_type().enumerate() {
        if r.eq(rtype) {
            let res = F::parse(pair, i)?;
            return Ok(res);
        }
    }
}
\end{lstlisting}

While we are visiting each node of the tree, the parser will update the compiler's context to track custom function declarations, structs, maps, and the program name. This context is a custom Rust object that provides the compiler with additional metadata about the Presto AST.

\begin{lstlisting}[breaklines]{rust}
struct Context {
    funcs: HashMap<String, FunctionDeclaration>,
    structs: HashMap<String, StructStatement>,
    mappings: HashMap<String, MappingStatement>,
    imports: HashSet<String>,
    constraints: HashSet<String>,
    keys: HashSet<String>,
    add_constraint: bool,
    program_name: String,
}
\end{lstlisting}

\subsubsection{Interpreter}

At this stage, the Interpreter will take the AST that we got from running our additional parsing logic as its only input. This is the stage that will analyze the computational workload of a given Presto file. Using similar logic to the \texttt{try\_parse} trait, we create a \texttt{try\_interpret} trait that will achieve the same objective of recursively traversing the AST.

\begin{lstlisting}[breaklines]{rust}
trait Interpret<F, R> where R: PrestoNode<R> {
    fn interpret(ctx: Context, r: R) -> Result<F>;
}

fn try_interpret<R, I>(ctx: Context, r: R) -> Result<I>
where
    R: PrestoNode<R>,
    I: Interpret<I, R>,
{
    I::interpret(ctx, r)
}

fn interpret_ast(ctx: Context, ast: PrestoFile) -> Result<HashMap<String, u64>> {
    let statements = ast.statements;
    let res = try_interpret::<Statements, StatementsInterpreter>(ctx, statements)
        .usage_table;
    Ok(res)
}
\end{lstlisting}

The Interpreter begins with the global scope of the Presto program and is looking for variable declarations that contain integer operations and specific Presto function calls, such as \texttt{sha256}, \texttt{mimc}, and \texttt{keccak256}. Every node will contain a \texttt{usage\_table} Hashmap and \texttt{uses\_storage} boolean as additional metadata for the zkSDK compiler. The \texttt{usage\_table} will record the type of operation and the number of times it appears throughout the AST. 

The Interpreter naively traverses all function declarations— regardless of whether they're actually used— in order to analyze each function's \texttt{usage\_table}. It will then track the number of times the function is called and multiply the corresponding \texttt{usage\_table} by that count. The Interpreter will also naively traverse conditionals without evaluating the statement.

Furthermore, the Interpreter will check for any access to storage. If storage is accessed, it will track the number of unique keys used in the program. In the Presto language, this only occurs when a mapping statement is called. This would flag the code generation section to properly handle the storage-related code. All of this information will be written into the Context object.

When handling conditionals, the Interpreter will naively 

\begin{lstlisting}[breaklines]{rust}
struct ExpressionInterpreter {
    usage_table: HashMap<String, u64>,
    uses_storage: bool,
}

impl Interpret<ExpressionInterpreter, Expression> for ExpressionInterpreter {
    fn interpret(ctx: Context, r: Expression) -> Result<ExpressionInterpreter> {
        let usage_table = HashMap::new();
        let uses_storage = false;

        match r {
            Expression::CustomFunctionCall(cfc) => {
                match cfc.name {
                    "sha256" | "sha3_keccak256" | "any-other-known-fn" => {
                        // increase usage in usage table
                    }
                    _ => {
                        // custom fn
                        // recurse into impl
                    }
                }
            }
            Expression::Expr(expr) => {
                // record expression evaluation
            },
            _ => (),
        }

        Ok(ExpressionInterpreter {
            usage_table,
            uses_storage,
        })
    }
}
\end{lstlisting}

\subsubsection{Dynamic Backend Selection Mechanism}

After the AST parser and interpreter has been executed, the zkSDK will gather the user preference struct and convert into a 1x3 matrix, representing proof generation time, proof verification cost, and hardware acceleration as the initial set of preferences.

\begin{lstlisting}[breaklines]{rust}
impl Preference {
    fn to_matrix(&self) -> Vector3<f64> {
        Vector3::new(
            self.proof_generation,
            self.proof_verification,
            self.hardware_acceleration,
        )
    }
}
\end{lstlisting}

The zkSDK will also convert the Interpreter results into a 1x11 matrix, where each row corresponds to the type of benchmark recorded, as well as normalize the benchmark results. With these inputs, the zkSDK performs matrix multiplication on the \texttt{usage\_table} and \texttt{normalized\_table} and selects the cell with the smallest value.

\begin{lstlisting}[breaklines]{rust}
fn dynamic_select(interpret: HashMap<String, u64>, p: &Preference) -> Backend {
    let usage = get_interpret_matrix(interpret);
    let normalized = get_table_normalized();
    let pref_vec = p.to_matrix();

    let scores = normalized * usage * pref_vec;

    // Find row with smallest value
    let row = scores
        .enumerate()
        .min_by(|(_, a), (_, b)| a.partial_cmp(b));

    // Find corresponding col
    let col = row
        .enumerate()
        .min_by(|(_, a), (_, b)| a.partial_cmp(b));

    match col {
        0 => Backend::RiscZero,
        1 => Backend::GnarkGroth16,
        2 => Backend::GnarkPlonk,
    }
}
\end{lstlisting}

Although the first iteration of the DBSM is inherently simple, we will describe a future research direction on developing a more sophisticated DBSM. Assuming the user preferences remain the same, we will modify the hardware acceleration and proof verification input format; replacing the current weighted representation, which will allow for more a granular specification. 

\begin{itemize}
  \item Hardware acceleration format: \texttt{[benchmark]-\newline[backend]-[device]:[num\_cores]}. The new hardware acceleration format allows users to specify a preference for a CPU or GPU device, along with the number of cores to utilize. The user can also use the \texttt{current} keyword, indicating a preference for using their local device. For example, if a user has a MiMC intense workload and wants to use a CPU with 10 cores, they can write it exactly as \texttt{MiMC-gnark-cpu:10}.
  \item Proof Verification format: \texttt{[curve]-[zk\_schema]- \allowbreak [vm]:[max\_gas]}. This new format would mean that the ZKP should be verifiable on the \texttt{vm} using the \texttt{zk\_ \allowbreak schema} on the elliptic curve, \texttt{curve}, and consuming at most \texttt{max\_gas} units. For example, \texttt{BN254-Groth16- \allowbreak EVM:400000} would mean that the ZKP should be verifiable within 400,000 gas on the Ethereum Virtual Machine (EVM) using Groth16 on the BN254 curve.
\end{itemize}

If a user doesn't care about a specific preference, the wildcard, \texttt{*}, can be used. Now, we present the extended DBSM algorithm:

\begin{figure*}[h]
\centering
\includegraphics[width=0.9\textwidth]{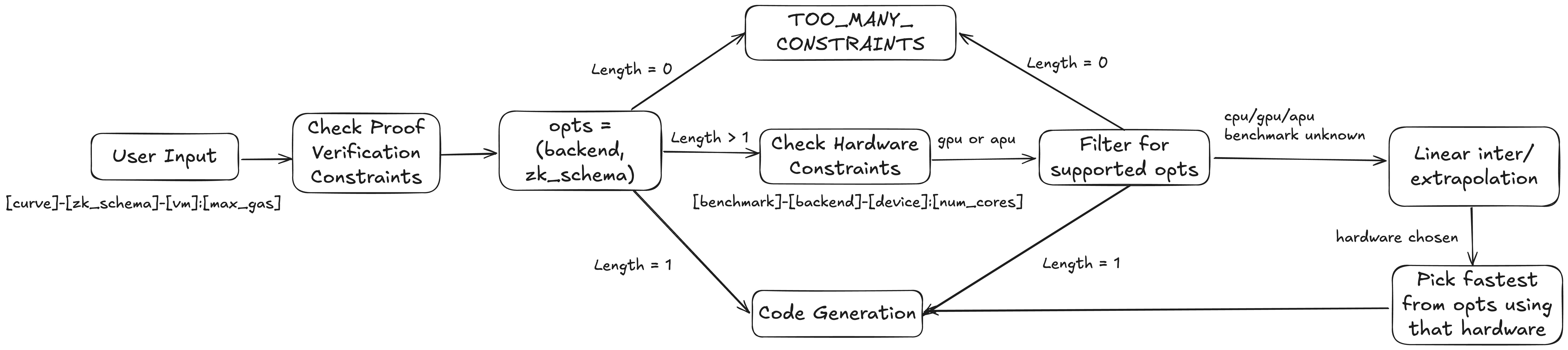}
\caption{zkSDK DBSM Extended Algorithm}
\label{fig:zksdk-dbsm-extended}
\end{figure*}

\begin{enumerate}
  \item Check proof verification constraint. For every backend and ZK schema that we have, keep the options that match the user input. If a backend doesn't support a ZK schema, this will be removed.  If \texttt{max\_gas} is provided, benchmark the program against the available options, and only keep the ones that are below the limit.
  \item If Step 1 returns no options, the DBSM exits with a Too Many Constraints message and suggests to modify the gas limit.
  \item If Step 1 returns one backend and ZK schema, the DBSM exits and immediately proceeds to the Codegen stage.
  \item If Step 1 returns multiple backend and ZK schema options, then consider the user's hardware acceleration preference:
  \vspace{0.5em}
    \begin{enumerate}
      \item If the user supplies \texttt{*}, return a baseline device (i.e., M1 Pro or a sufficient AWS machine).
      \item If the user supplies \texttt{current}, return the device's GPU cores. If the device does not have a GPU, return the number of CPU cores. If the device is an Apple Silicon, return \texttt{apu:M}. 
      \item If the user supplies \texttt{cpu} or \texttt{gpu}, return a baseline device.
      \item If the user supplies \texttt{cpu:N} or \texttt{gpu:M}, return the input.
    \end{enumerate}
  \item Let us consider the proof generation time using the selected hardware. It is important to note that we might not have the benchmark results for every hardware device. For the devices that we do, we can return this number and go to Step 10. 
  \item If the hardware device is a \texttt{gpu} or \texttt{apu}, return the backend and ZK schema options that support it.
  \item If there are no options left, the DBSM exits with a Too Many Constraints message and suggests to modify the hardware preference. 
  \item If there is one backend and ZK schema option left, the DBSM exits and immediately proceeds to the Codegen stage.
  \item If the selected hardware device does not have a benchmark result, then we will do linear interpolation. Suppose we have the results for two devices, X and Y. The DBSM selects device N, which we do not have the result for. If the proof generation time takes \texttt{T\_y} for device Y, and \texttt{T\_x} for device X, then for device N should be \texttt{T\_n = (T\_y-T\_x)/(y-x) * n + T\_x} , assuming that X < N < Y, in terms of hardware performance. If N < X and N > Y, then this becomes an extrapolation problem, which may yield higher error rates. Alternatively, the DBSM can modify its baseline to X or Y, respectively. 
  \item With the optimally selected hardware, the DBSM will return the fastest backend and ZK schema option.
\end{enumerate}

\subsubsection{Risc Zero Codegen}

Given that the risc0 zkVM Guest program supports Rust and is designed to be largely similar to writing a normal Rust program, this makes the translation from Presto fairly straightforward. Furthermore, since Presto is statically typed, we do not need to do any complicated type inferencing. For most of the language features, it is almost a one-to-one translation from Presto to Rust. This would include variable declarations, function definitions, binary and integer operations, byte manipulation, conditions, for-loops, and structs. These code translations would be generated in the Guest section of the Risc Zero codebase.

For Mappings, this would be analogous to BTreeMaps in Rust. Risc Zero makes a particular note about using \allowbreak BTreeMaps over Hashmaps \cite{risczeroopt} to improve performance in the zkVM. Another important consideration is that Mappings are not ephemeral storage. In the context of blockchains, Mappings can store sensitive information containing a wallet balance of a particular token. If the zkSDK Codegen generates the BTreeMap inside the Guest section, this would be problematic since the Host Executor creates a new instance of the Guest program each time causing the previous data to be lost! To address this, recall that one of responsibilities of the Interpreter is to update the Context to specify if storage was accessed. Along with the Host's ability to handle inputs and outputs of the Guest program execution, we can pass a BTreeMap in the Host section as input. Essentially, our solution mimics that of a modern blockchain scenario, where realistically, this would be replaced with a Merkle hash, and the Guest code would be updated to handle the deserialization. Due to time limitations, we could not implement this.

\begin{lstlisting}[breaklines]{rust}
  if !ctx.get_maps().is_empty() {
      ctx.get_maps().for_each(|m| {
          println!(
              "let {}: BTreeMap<{}, {}> = BTreeMap::new();",
              m.name,
              m.ty1,
              m.ty2
          );
      });
  }

  println!("let env = ExecutorEnv::builder()");
  ctx.get_maps().for_each(|m| {
      println!(".write(&{}).unwrap()", m.name);
  });
  println!(".build().unwrap();");
\end{lstlisting}

\subsubsection{Gnark Codegen}

Generating Gnark code is non-trivial given the nature of writing ZK circuits over traditional application code. However, since circuits primarily interacts with \texttt{frontend. \allowbreak Variable}'s, this makes the translation for variable declarations and return types straightforward, as we simply need to type annotate everything as a \texttt{frontend.Variable}. Furthermore, every variable defined at the global scope of the Presto program will be added as a \texttt{frontend.Variable} to the circuit. It is important to understand that a \texttt{frontend.Variable} is a constraint in the ZK circuit. Although we naively classify everything with this variable, it is important to not sacrifice the integrity of the circuit. We leave this optimization as an open area of future discussion. 

\begin{lstlisting}[breaklines]{go}
type tokenCircuit struct {
	IsPositiveBalance frontend.Variable
	Bal frontend.Variable
	Keys [2]frontend.Variable
	Values [2]frontend.Variable
}
\end{lstlisting}

For custom function definitions, it is a trivial translation with some added components. Every function contains a method receiver on the circuit, as well as passing the \newline \texttt{frontend.API} as an additional parameter to interface with the circuit. A special function called \texttt{Define} is created to contain the logic of the ZK circuit. This function body is populated with the global scope of the Presto program. 

\begin{lstlisting}[breaklines]{go}
func (circuit *tokenCircuit) Define(api frontend.API) error {
  circuit.mint_public(api, frontend.Variable("0xdeadbeef"), frontend.Variable(10))
  circuit.transfer_public(api, frontend.Variable("0xdeadbeef"), frontend.Variable("0xdeadbeef1"), frontend.Variable(1))

  var bal frontend.Variable = circuit.get_balance(api, "0xdeadbeef")
  var isPositiveBalance frontend.Variable = 0
  if api.Cmp(bal, 1) == 1 {
    isPositiveBalance = 1
  }

  circuit.IsPositiveBalance = isPositiveBalance
  circuit.Bal = bal
  return nil
}
\end{lstlisting}

When it comes to the Mapping statement, we cannot simply convert it into a Hashmap and type cast it as a \texttt{frontend.\allowbreak Variable}. This is where we rely on the Interpreter's results of finding the total unique keys that were accessed by the Presto program. With this information, the Gnark Codegen will created two \texttt{[]frontend.Variable} called \texttt{Keys} and \texttt{Values}, each with a length of the number of unique keys. By passing these lists as a public input to the witness, this will address the ephemeral issue that we discussed earlier.

\begin{lstlisting}[breaklines]{rust}
  println!("type {}Circuit struct {{", ast.name));
  ctx.get_constraints().for_each(constraint {
      println!(
          "{} frontend.Variable",
          constraint[0..1].to_uppercase() + &constraint[1..]
      );
  });
  println!(
      "Keys [{}]frontend.Variable",
      ctx.get_keys().len()
  );
  println!(
      "Values [{}]frontend.Variable",
      ctx.get_keys().len()
  );
  println!("}".to_string());
\end{lstlisting}

Other language features like structs, if-statements, for-loops, binary and integer operations are one-to-one translations, aside from the differences between Golang and Presto. The final part within the main body is to setup the proving logic. At the moment, the zkSDK Codegen only supports the BN254 curve on Groth16.

\begin{lstlisting}[breaklines]{go}
func main() {
  // ... global scope code ... // 

  // proving logic
  var circuit tokenCircuit
  witness := &tokenCircuit{IsPositiveBalance: 0, Bal: 0, Keys: keys, Values: values}
  privWit, err := frontend.NewWitness(witness, ecc.BN254.ScalarField())
  if err != nil { panic(err) }

  pubWit, err := frontend.NewWitness(witness, ecc.BN254.ScalarField(), frontend.PublicOnly())
  if err != nil { panic(err) }

  var ccs constraint.ConstraintSystem
  ccs, err = frontend.Compile(ecc.BN254.ScalarField(), r1cs.NewBuilder, &circuit)
  if err != nil { panic(err) }

  pk, vk, err := groth16.Setup(ccs)
  if err != nil { panic(err) }

  proof, err := groth16.Prove(ccs, pk, privWit)
  if err != nil { panic(err) }

  err = groth16.Verify(proof, vk, pubWit)
  if err != nil { panic(err) }
}
\end{lstlisting}
\section{Implementation}

Throughout the thesis, the following components were developed:

\begin{itemize}
  \item \textbf{Benchmarks:} This package contained the framework to benchmark proof generation times and generate the inputs for the proof verification in the Contracts package for the Risc Zero and Gnark backends. This was written in Rust and Go, respectively across 1200 lines of code.
  \item \textbf{Contracts:} The contracts package was developed to benchmark the gas usage of proof verification on \allowbreak Ethereum using \href{https://book.getfoundry.sh/}{Foundry} by taking the inputs from the Benchmarks package.
  \item \textbf{Compiler:} The compiler package was written in Rust and consisted of the AST Parser, Interpreter, and Codegen components. This was implemented with 3300 lines of code.
  \item \textbf{DBSM:} The selection algorithm was written in Rust and was responsible for handling the backend selection given the user preferences and computational profiling result.
\end{itemize}
\section{Evaluation}

\subsection{Setup}

The experimental environment consisted of the following technologies and tools:

\begin{itemize}
  \item \textbf{Macbook M1 Pro:} With 32GB of RAM and running MacOS v14.6.1
  \item \textbf{Rust version:} v1.84.0
  \item \textbf{Go version:} v1.23.4
  \item \textbf{Risc Zero versions:} risc0-zkvm v1.2.0 and risc0-build v1.2.0. To generate Groth16 proofs, we used Bonsai, Risc Zero's remote proving service.
  \item \textbf{Gnark version:} v0.11.0
  \item \textbf{Solidity testing:} We used Foundry, a popular, and standard for writing smart contracts on Ethereum. It uses forge v0.2.0.
  \item \textbf{Target bench time:} 60 seconds
\end{itemize}

\subsection{Results and Analysis}

The primary performance metric used in this evaluation is the proof generation time. This is the time that it takes to set up the proving system and generate the ZK proof. We ran this on two real-world examples, and the zkSDK was able to optimally select the appropriate ZK backend and generate the corresponding code. 

\subsubsection{Token Example}

This workload enables users to mint a token and then transfer these tokens to other recipients. It resembles the real-world example of issuing and transferring ERC20 tokens on the Ethereum blockchain. 

\begin{lstlisting}[breaklines]{python}
program token {
    mapping account: (owner pubkey => amount u64);

    struct Token {
        secretkey owner;
        u64 amount;
    }

    func mint(r0: pubkey, r1: u64) -> None {
        let r2: u64 = get_balance(r0);
        let r3: u64 = r1 + r2;
        set_balance(r0, r3);
        return;
    }

    /*
     * r0: Input the token send.
     * r1: Input the token receiver.
     * r2: Input the token amount.
     */ 
    func transfer(r0: pubkey, r1: pubkey, r2: u64) -> None {
        let r3: u64 = get_balance(r0);
        let r4: u64 = r3 - r2;
        
        // Decrements account[r0] by r2.
        set_balance(r0, r4);

        let r5: u64 = get_balance(r1);
        let r6: u64 = r5 + r2;
        
        // Increments account[r1] by r2.
        set_balance(r1, r6);

        return;
    }

    mint_(0xdeadbeef, 10);
    transfer(0xdeadbeef, 0xdeadbeef1, 1);

    let bal: u64 = get_balance(0xdeadbeef);
    let isPositiveBalance: bool = false;
    if (bal > 1) {
        isPositiveBalance = true;
    }
}
\end{lstlisting}

It is important to highlight that due to time limitations, we could not benchmark the memory access across both backends. Therefore, we made the assumption that memory access is relatively cheap. Upon inspection, this code is primarily integer operation heavy, as it consists of updating the appropriate balances. If we were to run this Presto code in both Risc Zero and Gnark, this would be the proof verification times:

\begin{table}[h]
\centering
\begin{tabular}{lr}
\toprule
Gnark & Risc Zero \\
\midrule
109.038ms $\pm$ 7.441ms & 1590ms $\pm$ 70ms \\
\bottomrule
\end{tabular}
\end{table}

With a user preference of favouring proof generation time, when we compile the Presto program with the zkSDK, the Interpreter returns a profile result indicating the specific number of integer operations. Therefore, knowing our benchmark results, we can see that the zkSDK DBSM properly selects the Gnark backend, as desired. 

\begin{lstlisting}[breaklines]{bash}
cargo run -- ../examples/token/token.presto 1 0 0

Interpreter Result:
{"int_ops": 6, "get_balance": 7, "set_balance": 6}

Backend Selection:
GnarkGroth16
\end{lstlisting}

\subsubsection{Merkle Tree Example}

The second example is constructing a Merkle Tree in Presto. This resembles a workload similar in blockchain applications where cryptographic hashes are stored in a hierarchical structure for efficient storage and retrieval of historical data. 

\begin{lstlisting}[breaklines]{python}
program merkle_tree {
    func build_hash(leaf1: bytes, leaf2: bytes) -> bytes {
        let left: bytes = sha256(leaf1);
        let right: bytes = sha256(leaf2);

        let combined: bytes = extend_vec(left, right);
        return sha256(combined);
    }

    func build_merkle_root(leaf1: bytes, leaf2: bytes, leaf3: bytes, leaf4: bytes) -> bytes {
        let n1: bytes = build_hash(leaf1, leaf2);
        let n2: bytes = build_hash(leaf3, leaf4);

        return build_hash(n1, n2);
    }

    let leaf1: bytes = 0x12345;
    let leaf2: bytes = 0x67890;
    let leaf3: bytes = 0x11111;
    let leaf4: bytes = 0x22222;

    let merkle_root: bytes = build_merkle_root(leaf1, leaf2, leaf3, leaf4);
}
\end{lstlisting}

Constructing this Merkle Tree was done using SHA-256 hashes. If we were to run this example on Risc Zero and Gnark, then we would expect to see Risc Zero perform better at computing SHA-256 hashes:

\begin{table}[h]
\centering
\begin{tabular}{lr}
\toprule
Gnark & Risc Zero \\
\midrule
3401ms $\pm$ 239ms & 2700ms $\pm$ 60ms \\
\bottomrule
\end{tabular}
\end{table}

With a user preference for favoring proof generation time, when we compile this with the zkSDK, the Interpreter returns a profile result indicating the specific number of SHA-256 operations. Therefore, knowing our benchmark results, we can see that the zkSDK DBSM properly selects the Risc Zero backend.  

\begin{lstlisting}[breaklines]{bash}
cargo run -- ../examples/merkle_tree/merkle_tree.presto 1 0 0

Interpreter Result:
{"sha256": 17}

Backend Selection:
Risc Zero
\end{lstlisting}
\section{Related Work}

There has been a lot of research and development in making ZKP development as seamless and accessible as possible. Given the infancy of ZKP development, there remains a lot of further engineering optimizations. As a result, many of these advances are industry-driven, and in this section, we will cover key projects and their strengths, trade-offs, and relevance to zkSDK.

Several projects introduced high-level programming languages like Leo by Aleo \cite{aleo} and o1js \cite{o1js} by O(1) Labs to make ZKP development accessible to developers who already know a general-purpose programming language. These languages aim to abstract the underlying ZK systems and provide a normal developer experience like writing standard application code. However, these languages were designed to be used only on their respective protocols, limiting developers to easily move their ZK application to another backend/platform.

On the other hand, there are projects that support multiple ZK backends. ZoKrates \cite{zokrates} offers a high-level language along with supporting different elliptic curves, schemes, and two backends: Bellman and Ark. However, this is a manual backend selection that the developer needs to specify. In contrast, Z\O\ \cite{zo} supports two backends and optimally selects the appropriate option through a cost-model analysis. Users can interface with this using the C\# programming language. Although the two backends, ZQL and Pinocchio, are outdated, they introduced the idea of dynamic selection.

As we can see, none of the projects employ the idea of utilizing a high-level programming language that isn't tailored towards any execution environment, as well as automatically selecting the ideal ZK backend based on traces of the program.
\section{Conclusion}

In this thesis, we introduced zkSDK, a novel framework that leverages Presto, a custom-built programming language that abstracts the complexities of choosing ZK-proving backends. By dynamically selecting the optimal option based on user-defined criteria and computational traces, users would not have to worry about learning different ZK backends or suffer from performance, as the zkSDK can support various backends in the future.

This work opens the door for future research to improve the capability of the Interpreter to intelligently divide the Presto program into segments to prove on different ZK backends based on its performance strengths. Although the tradeoffs and performance gains are unknown at this time, it remains an interesting and open area of research.

\bibliographystyle{ACM-Reference-Format}
\bibliography{refs}


\begin{thebibliography}{14}


\ifx \showCODEN    \undefined \def \showCODEN     #1{\unskip}     \fi
\ifx \showDOI      \undefined \def \showDOI       #1{#1}\fi
\ifx \showISBNx    \undefined \def \showISBNx     #1{\unskip}     \fi
\ifx \showISBNxiii \undefined \def \showISBNxiii  #1{\unskip}     \fi
\ifx \showISSN     \undefined \def \showISSN      #1{\unskip}     \fi
\ifx \showLCCN     \undefined \def \showLCCN      #1{\unskip}     \fi
\ifx \shownote     \undefined \def \shownote      #1{#1}          \fi
\ifx \showarticletitle \undefined \def \showarticletitle #1{#1}   \fi
\ifx \showURL      \undefined \def \showURL       {\relax}        \fi
\providecommand\bibfield[2]{#2}
\providecommand\bibinfo[2]{#2}
\providecommand\natexlab[1]{#1}
\providecommand\showeprint[2][]{arXiv:#2}

\bibitem[\protect\citeauthoryear{Aleo}{Aleo}{2025}]%
        {aleo}
\bibfield{author}{\bibinfo{person}{Aleo}.} \bibinfo{year}{2025}\natexlab{}.
\newblock \bibinfo{title}{The Leo Programming Language}.
\newblock \bibinfo{howpublished}{\url{https://docs.leo-lang.org/leo/}}.   (\bibinfo{year}{2025}).
\newblock
\newblock
\shownote{Accessed: 2025-03-30.}


\bibitem[\protect\citeauthoryear{{Chain Link}}{{Chain Link}}{2023a}]%
        {chainlinkzk}
\bibfield{author}{\bibinfo{person}{{Chain Link}}.} \bibinfo{year}{2023}\natexlab{a}.
\newblock \bibinfo{title}{Zero-Knowledge Proofs: What are ZK Proofs?}
\newblock \bibinfo{howpublished}{\url{https://chain.link/education/zero-knowledge-proof-zkp}}.   (\bibinfo{year}{2023}).
\newblock
\newblock
\shownote{Accessed: 2025-03-23.}


\bibitem[\protect\citeauthoryear{{Chain Link}}{{Chain Link}}{2023b}]%
        {chainlinksnark}
\bibfield{author}{\bibinfo{person}{{Chain Link}}.} \bibinfo{year}{2023}\natexlab{b}.
\newblock \bibinfo{title}{ZK-SNARKs vs. ZK-STARKs: A Comparative Analysis}.
\newblock \bibinfo{howpublished}{\url{https://chain.link/education-hub/zk-snarks-vs-zk-starks}}.   (\bibinfo{year}{2023}).
\newblock
\newblock
\shownote{Accessed: 2025-03-23.}


\bibitem[\protect\citeauthoryear{{Coinmonks}}{{Coinmonks}}{2023}]%
        {plonk}
\bibfield{author}{\bibinfo{person}{{Coinmonks}}.} \bibinfo{year}{2023}\natexlab{}.
\newblock \bibinfo{title}{Under the Hood of zk-SNARKs: PLONK Protocol - Part 1}.
\newblock \bibinfo{howpublished}{\url{https://medium.com/coinmonks/under-the-hood-of-zksnarks-plonk-protocol-part-1-34bc406d8303}}.   (\bibinfo{year}{2023}).
\newblock
\newblock
\shownote{Accessed: 2025-03-23.}


\bibitem[\protect\citeauthoryear{{Consensys}}{{Consensys}}{2023}]%
        {gnarkwit}
\bibfield{author}{\bibinfo{person}{{Consensys}}.} \bibinfo{year}{2023}\natexlab{}.
\newblock \bibinfo{title}{Gnark Documentation: Serialize}.
\newblock \bibinfo{howpublished}{\url{https://docs.gnark.consensys.io/HowTo/serialize}}.   (\bibinfo{year}{2023}).
\newblock
\newblock
\shownote{Accessed: 2025-03-23.}


\bibitem[\protect\citeauthoryear{Labs}{Labs}{2025}]%
        {o1js}
\bibfield{author}{\bibinfo{person}{O1 Labs}.} \bibinfo{year}{2025}\natexlab{}.
\newblock \bibinfo{title}{TypeScript framework for zk-SNARKs and zkApps}.
\newblock \bibinfo{howpublished}{\url{https://github.com/o1-labs/o1js}}.   (\bibinfo{year}{2025}).
\newblock
\newblock
\shownote{Accessed: 2025-03-30.}


\bibitem[\protect\citeauthoryear{Matthew~Fredrikson}{Matthew~Fredrikson}{2014}]%
        {zo}
\bibfield{author}{\bibinfo{person}{Benjamin~Livshits Matthew~Fredrikson}.} \bibinfo{year}{2014}\natexlab{}.
\newblock \bibinfo{title}{An optimizing distributed zero knowledge compiler}.
\newblock \bibinfo{howpublished}{\url{https://www.usenix.org/conference/usenixsecurity14/technical-sessions/presentation/fredrikson}}.   (\bibinfo{year}{2014}).
\newblock
\newblock
\shownote{Accessed: 2025-03-30.}


\bibitem[\protect\citeauthoryear{Rezaei}{Rezaei}{2023}]%
        {riftv2}
\bibfield{author}{\bibinfo{person}{Amin Rezaei}.} \bibinfo{year}{2023}\natexlab{}.
\newblock \bibinfo{title}{Rift-v2: A framework for ZK development}.
\newblock \bibinfo{howpublished}{\url{https://github.com/AminRezaei0x443/rift-v2}}.   (\bibinfo{year}{2023}).
\newblock
\newblock
\shownote{Accessed: 2025-03-23.}


\bibitem[\protect\citeauthoryear{{Risc Zero}}{{Risc Zero}}{2023a}]%
        {risczeroproving}
\bibfield{author}{\bibinfo{person}{{Risc Zero}}.} \bibinfo{year}{2023}\natexlab{a}.
\newblock \bibinfo{title}{Risc Zero Local Proving: Proving Hardware}.
\newblock \bibinfo{howpublished}{\url{https://dev.risczero.com/api/generating-proofs/local-proving##proving-hardware}}.   (\bibinfo{year}{2023}).
\newblock
\newblock
\shownote{Accessed: 2025-03-23.}


\bibitem[\protect\citeauthoryear{{Risc Zero}}{{Risc Zero}}{2023b}]%
        {risczeroopt}
\bibfield{author}{\bibinfo{person}{{Risc Zero}}.} \bibinfo{year}{2023}\natexlab{b}.
\newblock \bibinfo{title}{Risc Zero Optimization: TLDR and Quick Wins}.
\newblock \bibinfo{howpublished}{\url{https://dev.risczero.com/api/zkvm/optimization##tldr-and-quick-wins}}.   (\bibinfo{year}{2023}).
\newblock
\newblock
\shownote{Accessed: 2025-03-23.}


\bibitem[\protect\citeauthoryear{{Risc Zero}}{{Risc Zero}}{2023c}]%
        {risczeroreceipt}
\bibfield{author}{\bibinfo{person}{{Risc Zero}}.} \bibinfo{year}{2023}\natexlab{c}.
\newblock \bibinfo{title}{Risc Zero Terminology: Receipt}.
\newblock \bibinfo{howpublished}{\url{https://dev.risczero.com/terminology##receipt}}.   (\bibinfo{year}{2023}).
\newblock
\newblock
\shownote{Accessed: 2025-03-23.}


\bibitem[\protect\citeauthoryear{{Risc Zero}}{{Risc Zero}}{2023d}]%
        {risczero}
\bibfield{author}{\bibinfo{person}{{Risc Zero}}.} \bibinfo{year}{2023}\natexlab{d}.
\newblock \bibinfo{title}{Risc Zero Use Cases}.
\newblock \bibinfo{howpublished}{\url{https://dev.risczero.com/api/use-cases}}.   (\bibinfo{year}{2023}).
\newblock
\newblock
\shownote{Accessed: 2025-03-23.}


\bibitem[\protect\citeauthoryear{{Risc Zero}}{{Risc Zero}}{2023e}]%
        {risczerozvm}
\bibfield{author}{\bibinfo{person}{{Risc Zero}}.} \bibinfo{year}{2023}\natexlab{e}.
\newblock \bibinfo{title}{Risc Zero zkVM API}.
\newblock \bibinfo{howpublished}{\url{https://dev.risczero.com/api/zkvm/}}.   (\bibinfo{year}{2023}).
\newblock
\newblock
\shownote{Accessed: 2025-03-23.}


\bibitem[\protect\citeauthoryear{ZoKrates}{ZoKrates}{2025}]%
        {zokrates}
\bibfield{author}{\bibinfo{person}{ZoKrates}.} \bibinfo{year}{2025}\natexlab{}.
\newblock \bibinfo{title}{ZoKrates: Introduction}.
\newblock \bibinfo{howpublished}{\url{https://zokrates.github.io/}}.   (\bibinfo{year}{2025}).
\newblock
\newblock
\shownote{Accessed: 2025-03-30.}


\end{thebibliography}
\end{document}